\documentclass[showpacs,preprintnumbers,amsmath,amssymb]{revtex4}
\usepackage{amsmath,amssymb,graphics,epsfig,subfigure}
\usepackage{color}

\begin{document}
\newcommand {\nn} {\nonumber}
\renewcommand{\baselinestretch}{1.3}

\title{Characteristic precessions of spherical orbit around a rotating braneworld black hole}

\author{Hui-Min Wang$^{a}$,
	    Kai Liao$^{a}$\footnote{liaokai@whu.edu.cn, corresponding author},
	    Shao-Wen Wei$^{b,c,d}$ \footnote{weishw@lzu.edu.cn, corresponding author}}

\affiliation{
	$^{a}$School of Physics and Technology, Wuhan University, Wuhan 430072, China,\\
	$^{b}$Lanzhou Center for Theoretical Physics, Key Laboratory of Theoretical Physics of Gansu Province, School of Physical Science and Technology, Lanzhou University, Lanzhou 730000, China,\\
	$^{c}$Institute of Theoretical Physics $\&$ Research Center of Gravitation,Lanzhou University, Lanzhou 730000, China,\\
	$^{d}$ Gansu Provincial Research Center for Basic Disciplines of Quantum Physics}

\begin{abstract}
	We study the orbital dynamics and relativistic precession effects in the spacetime of rotating braneworld black holes within the Randall-Sundrum framework. For test particles on spherical orbits, we analyze three conserved quantities—energy, angular momentum, and Carter constant—and examine how the innermost stable spherical orbit depends on the tidal charge and orbital inclination. Compared to Kerr black holes, braneworld corrections significantly modify both nodal and periastron precession frequencies: positive tidal charges suppress precession rates, while negative charges enhance them. For stationary gyroscopes, we calculate the Lense-Thirring precession frequency and demonstrate its sensitivity to the tidal charge, black hole spin, and gyroscope orientation. Our results show that a positive tidal charge weakens frame-dragging effects even as it enhances gravitational attraction—offering a distinctive signature of extra-dimensional gravity. These results have important implications for astrophysical observations, including accretion disk behavior, stellar orbital dynamics, and gravitational wave detection. The modified orbital and gyroscopic precession provide new ways to test braneworld gravity in strong-field regimes.
\end{abstract}

\maketitle

\section{Introduction}
\label{sec:Introduction}

The Event Horizon Telescope (EHT) collaboration has provided direct evidence for one of the key predictions of general relativity (GR)—the existence of black holes—by releasing images of the intensity and polarization structures of the supermassive black holes (SMBHs) in M87* \cite{EventHorizonTelescope:2019dse,EventHorizonTelescope:2019ggy,EventHorizonTelescope:2021bee,EventHorizonTelescope:2024dhe} and Sgr A* \cite{EventHorizonTelescope:2022wkp,EventHorizonTelescope:2022xqj,EventHorizonTelescope:2024hpu}. These observations have yielded detailed measurements of black hole properties—size and the nature of the event horizon. They have also revealed the geometric structure of the surrounding spacetime, illustrating how black holes distort the fabric of spacetime. Such information is crucial for testing GR in extreme gravitational environments and refining theoretical models of black hole behavior. The constraints derived from these observations help narrow the range of possible gravitational theories, offering potential insights into the fundamental nature of gravity and spacetime. These achievements represent a significant advance in our understanding of black holes and GR.

Recently, the EHT has posed four fundamental science questions \cite{EventHorizonTelescope:2024whi}, the first of which concerns the origin of relativistic jets from SMBHs. In response, an observational goal has been proposed: to model the dynamical link between the central SMBH in M87 and its relativistic jet. Notably, in Ref. \cite{Cui:2023uyb}, Cui et al. pointed out that M87 provides a unique opportunity to study the connection between SMBHs and relativistic jets. Through analysis of the central radio source, they identified an asymmetric ring structure consistent with GR predictions. Moreover, they found that the angular orientation of the black hole’s jet varies periodically, with a period of roughly 11 years—leading to the hypothesis that this behavior is caused by Lense–Thirring precession from the central rotating black hole. This observation has motivated several theoretical efforts to model spherical orbits responsible for jet precession in M87, taking into account spin-induced dynamics in Kerr \cite{Wei:2024cti} and other black hole spacetimes \cite{Chen:2024aaa,Meng:2024gcg}.

The Lense–Thirring effect is a prediction of GR that describes how a massive rotating object, such as a black hole, drags spacetime around it, causing nearby objects to precess \cite{Lense:1918zz}. Empirical observation of this effect is vital for validating GR. For instance, Gravity Probe B, launched by NASA in 2004, successfully measured the Lense–Thirring precession of gyroscopes in Earth's orbit, thereby confirming the effect’s existence \cite{Everitt:2011hp}. In parallel, theoretical studies have analyzed Lense–Thirring precession in various spacetime backgrounds \cite{Wu:2023wld,AlZahrani:2023xix,Zhen:2025nah,Iorio:2024eey}. In the realm of orbital mechanics pertaining to satellites and space probes, the incorporation of the Lense-Thirring effect is imperative to ensure the accuracy of orbital predictions. In gravitational wave astronomy, Lense-Thirring effect plays a key role in determining the characteristics and evolution of compact binary systems, influencing the waveform signatures detected by observatories such as LIGO and Virgo. Thus, this effect is not just a theoretical construction but a critical component of both astrophysics and space technology. Accurate measurement of Lense–Thirring effect enhances our understanding of gravity and contributes to technological advances.

When the angular momentum of a black hole is misaligned with that of an orbiting particle, complex dynamical interactions arise as a result of precession torque. In the 1970s, Bardeen and Petterson investigated the dynamics of accretion disks around Kerr black holes, focusing on the Lense–Thirring effect in tilted disks \cite{Bardeen:1975zz}. Through rigorous mathematical modeling and theoretical analysis, they examined how this relativistic effect alters the stability and structure of accretion disks, as well as its ability to modify the orbital trajectories of embedded disk material. Their study demonstrated that the Lense-Thirring effect can drive precession in the spin axis of a test gyroscope—changing its orientation while preserving a fixed orbital path. The precession rate depends on several key parameters: the black hole's spin, the observer's radial distance from the black hole, and the observer's inclination relative to the black hole's rotational plane.

Building on this, Petterson further investigated how non-axisymmetric forces—particularly those associated with precession dynamics—affect thin accretion disks \cite{Petterson1977}. His theoretical framework predicted that disk annuli at different radii could maintain distinct orbital planes under such forces. By deriving the fundamental equations governing these warped disk configurations, Petterson established critical theoretical foundations for understanding accretion disk behavior in the strong gravitational fields of compact objects.

Subsequent research expanded the study of twisted disks to larger scales. Ref. \cite{Ostriker1989} analyzed how infalling extragalactic material might warp galactic disks, proposing a steadily varying axisymmetric potential. This model explains warps in outer galactic regions and changes in the Milky Way’s nuclear disk orientation.

Another major scientific question raised by the EHT in its Mid-Range Science Objectives is: \emph{What are the characteristics of the massive compact objects in galactic nuclei?} \cite{EventHorizonTelescope:2024whi} To address this, the EHT will measure features such as emission rings and apparent shadows of M87* and Sgr A*. Importantly, they pointed out that these features will depend sensitively on the underlying spacetime. While GR has passed all experimental tests so far, theoretical challenges like black hole singularities and the information paradox, along with cosmological puzzles such as dark matter and dark energy, suggest that GR might require modification. This motivates the exploration of alternative gravity theories—particularly those involving extra dimensions. High-dimensional gravity can be tested through detailed analysis of gravitational waves from black hole mergers \cite{LIGOScientific:2017bnn,Pardo:2018ipy}.

In this context, braneworld models \cite{Maartens:2010ar} have emerged as a leading framework, modeling our observable universe as a four-dimensional brane embedded in a higher-dimensional bulk. Derived from string theory and extra-dimensional gravity, these models offer new perspectives on the hierarchy problem and dark sector physics \cite{Brax:2004xh}, and naturally connect high-energy physics with cosmology through their extended spacetime structure.

One striking prediction of the braneworld model is the existence of braneworld black holes, which differ significantly from those predicted by GR \cite{Tanahashi:2011xx}. These black holes may feature altered event horizons, thermodynamics, and gravitational wave signatures due to the influence of extra-dimensional effects and the interaction between the brane and bulk. These features make braneworld black holes an exceptional testing ground for theories of gravity and high-dimensional physics.

In the past two decades, significant progress has been made in understanding braneworld black holes. Theoretical models have explored their formation mechanisms, stability, and thermodynamic properties. Models such as the Randall-Sundrum black hole and the Dvali-Gabadadze-Porrati black hole have been extensively studied. These solutions reveal that the presence of extra dimensions can influence black hole dynamics, such as Hawking radiation and accretion processes. In terms of observation, efforts to constrain braneworld scenarios have utilized data from gravitational wave detectors \cite{Yu:2016tar}, such as LIGO and Virgo, and telescopic observations of black hole shadows \cite{Vagnozzi:2019apd,Banerjee:2019nnj,Hou:2021okc}, such as those captured by the EHT. While experimental evidence for extra dimensions remains elusive, improving observational constraints continue to exclude significant regions of the braneworld parameter space.

Despite these advances, challenges remain—particularly in modeling brane-bulk interactions and distinguishing braneworld effects from other deviations in gravity theories.

Thus, black holes in the braneworld scenario may exhibit novel gravitational effects, including modifications to Lense–Thirring precession. Extra dimensions may introduce new forces or fields, altering how spin and orbital motion interact. Furthermore, gravity may propagate differently in higher dimensions, affecting the behavior of test particles and observables such as precession rates. Dadhich et al. \cite{Dadhich:2000am} provided a static, spherically symmetric vacuum brane solution in the Randall-Sundrum scenario \cite{Randall:1999ee,Randall:1999vf}, which serves as a foundational result for understanding black hole solutions in this framework. Furthermore, Aliev and Gumrukcuoglu \cite{Aliev:2005bi} extended these results to charged rotating black holes, highlighting the complexities introduced by rotation and charge in braneworld models.

These considerations motivate a detailed exploration of how extra-dimensional effects influence particle dynamics and relativistic precession, as we present in this work. In this paper, we perform a detailed investigation of particle dynamics and precession effects in the spacetime of rotating braneworld black holes within the Randall–Sundrum model. In particular, we focus on spherical orbits that deviate from the equatorial plane and explore how the presence of a tidal charge—a distinctive feature of braneworld gravity—modifies key aspects of orbital motion. Our analysis includes the behavior of conserved quantities, the location of the innermost stable spherical orbit (ISSO), and two types of relativistic precession: nodal (related to frame-dragging) and periastron (related to spacetime curvature). In addition, we study the Lense–Thirring precession of stationary gyroscopes, revealing how extra-dimensional effects influence spin precession rates. These results not only deepen our theoretical understanding of braneworld black holes, but also provide concrete observational predictions for testing extra-dimensional gravity using electromagnetic and gravitational wave measurements.

The structure of this article is as follows: In Section \ref{sec:braneworldbh}, we study test particle dynamics around rotating braneworld black holes, examining conserved quantities for spherical orbits and investigating the ISSO behavior as functions of the tidal charge and tilt angle. The nodal and periastron precession frequencies and their dependence on the braneworld parameters are analyzed in Section \ref{sec:particleprecession}. In Section \ref{sec:precessionlt}, we focus on the Lense-Thirring effect for stationary gyroscopes. Finally, we summarize our results and discuss their relevance for astrophysical observations and future experiments in Section \ref{sec:summary}.

\section{Particle motion around the rotating braneworld black hole}
\label{sec:braneworldbh}

At the beginning of this section, we briefly review the gravitational framework and the black hole solution adopted in our study. We consider the rotating black hole solution obtained in the Randall-Sundrum braneworld scenario, where our four-dimensional universe is modeled as a brane embedded in a higher-dimensional bulk spacetime. This model extends GR by introducing extra terms in the effective gravitational field equations—manifestations of the extra dimension. A key feature is the tidal charge $b$, which reflects the influence of the bulk geometry on brane-localized gravity and may take either positive or negative values.

The rotating black hole solution in this scenario, expressed in Boyer–Lindquist coordinates, is given by \cite{Aliev:2005bi}:
\begin{eqnarray}
	ds^2=&-&\biggl(1-\frac{2Mr-b}{\rho^2}\biggr)dt^2-2a\sin^2\theta\biggl(\frac{2Mr-b}{\rho^2}\biggr)dtd\phi\\
	&+&\sin^2\theta\biggl[\frac{(r^2+a^2)^2-\Delta a^2\sin^2\theta}{\rho^2}\biggr]d\phi^2+\frac{\rho^2}{\Delta} dr^2+\rho^2 d\theta^2,
\end{eqnarray}
where the metric functions are given by
\begin{eqnarray}
	\Delta&=&r^2-2Mr+a^2+b,\\
	\rho^2&=&r^2+a^2\cos^2\theta.
\end{eqnarray}
Here, $M$ and $a$ are the mass and spin (angular momentum per unit mass) of the black hole. The dimensionless parameter $b$ quantifies the tidal charge, a distinctive feature of the braneworld scenario that encodes the gravitational effects of the extra dimension on the four-dimensional brane. Unlike the electric charge in the Kerr–Newman solution, which appears as a squared term in the metric, the tidal charge $b$ can be either positive or negative, reflecting distinct bulk-induced effects.

A positive tidal charge $b>0$ enhances the gravitational pull near the black hole, leading to increased spacetime curvature and stronger confinement of geodesics—effectively making the black hole more compact. This causes the event horizon and the ISSO to move inward compared to the Kerr case. Conversely, a negative tidal charge introduces a repulsive effect, weakening the gravitational attraction and expanding the horizon and ISSO. These variations can impact the structure of the ergosphere, orbital stability, and observable phenomena such as quasi-periodic oscillations (QPOs) or jet formation.

The tidal charge $b$ uniquely characterizes braneworld effects, altering black hole geometry and dynamics. These modifications are observable in gravitational wave signals and black hole shadow images.

The location of the event horizon is determined by the condition $\Delta=0$, yielding:
\begin{eqnarray}
r_{\pm}=M\pm\sqrt{M^2-a^2-b}.
\end{eqnarray}
For $b<M^2-a^2$, the black hole has two distinct horizons: an outer horizon $r_{+}$ and an inner horizon $r_{-}$. In the extremal case where $b=M^2-a^2$, these two horizons coincide. If $b>M^2-a^2$, the solution represents a naked singularity rather than a black hole. These constraints have direct implications for astrophysical black holes. For example, in the case of M87* with a spin parameter of $a=0.9375M$ \cite{Cui:2023uyb}, we obtain the upper bound $b\leq 0.1211M^2$ for a physically acceptable black hole. The outer horizon $r_{+}$ is of particular importance as it marks the boundary beyond which no particle or light can escape the black hole's gravitational pull. This horizon acts as a one-way membrane, encapsulating the region where the black hole's intense gravity dominates. Outside the event horizon lies the ergosphere, a region defined by the condition $g_{tt}=0$. The outer boundary of the ergosphere is given by:
\begin{eqnarray}
	r_{\rm E}=M+\sqrt{M^2-a^2\cos^2\theta-b}.
\end{eqnarray}
On the equatorial plane, the ergosphere spans from the outer horizon $r_+$ to $r_{\rm E}$. Within this region, the dragging of inertial frames becomes so intense that no observer can remain stationary relative to infinity.

These geometric features strongly influence particle motion. Inside the ergosphere, co-rotating particles experience reduced effective gravity, allowing prograde ISSOs to approach closer to the black hole than retrograde ones. The tidal charge $b$ thus becomes a crucial parameter shaping the observable dynamics near the black hole.

Particle motion near a rotating braneworld black hole is governed by the geodesic equations derived from the Hamilton-Jacobi formalism. These trajectories are shaped by the black hole's strong gravitational field, which is further modified by the tidal charge $b$—a hallmark of extra-dimensional influence.

We consider two types of orbits: equatorial orbits, which lie in the plane perpendicular to the black hole's spin axis, and spherical (non-equatorial) orbits, which have non-zero Carter constants. Equatorial orbits are especially relevant for modeling accretion disks, while spherical orbits provide a more general framework for analyzing the motion of particles and matter in the black hole’s spacetime.

Using the Hamilton–Jacobi method, the geodesic equations for a test particle in the rotating braneworld black hole spacetime take the following form:
\begin{eqnarray}
	\rho^{2}\frac{dt}{d\lambda}&=&a(L-aE\sin^2\theta)+\frac{r^2+a^2}{\Delta}[(r^2+a^2)E-aL],\\
	\rho^{2}\frac{d\phi}{d\lambda}&=&\frac{L}{\sin^2\theta}-aE+\frac{a}{\Delta}[(r^2+a^2)E-aL],\\
	\label{Requation}
	\rho^{2}\frac{dr}{d\lambda}&=&\pm\sqrt{{\cal R}(r)},\\
	\rho^{2}\frac{d\theta}{d\lambda}&=&\pm\sqrt{\Theta(\theta)},
\end{eqnarray}
where
\begin{eqnarray}
	{\cal R}(r)&=&[(r^2+a^2)E-aL]^2-\Delta[r^2+{\cal K}+(L-aE)^2],\\
	\quad\Theta(\theta)&=&{\cal K} +\cos^2\theta\biggl(a^2E^2-\frac{L^2}{\sin^2\theta}-a^2\biggr).
\end{eqnarray}
Parameter $\lambda$ is the affine parameter along the geodesics, and $\cal K$ is the Carter constant, which reflects the motion in the polar direction. Energy $E$ and angular momentum $L$ constitute the canonical conserved quantities 
associated with the timelike and axial Killing vectors of the spacetime, respectively, dictating the particle's orbital 
dynamics.

\subsection{Tilted bound circular orbits}

To analyze off-equatorial orbits, we introduce the tilt angle $\zeta$, which characterizes the inclination of the orbital plane relative to the equatorial plane $\theta=\frac{\pi}{2}$. Specifically, the particle's orbital trajectory is confined between $\frac{\pi}{2}-\zeta$ and $\frac{\pi}{2}+\zeta$, allowing us to quantify deviations from equatorial symmetry. By varying $\zeta$, we can explore how tilt angles influence the particle's orbit, providing deeper insight into the effects of the black hole's gravitational field in the braneworld context. Based on the particle's motion in the angular direction, we can express the Carter constant ${\cal K}$ in terms of the tilt angle $\zeta$:
\begin{eqnarray}
	{\cal K}=a^2\sin^2\zeta(1-E^2)+L^2\tan^2\zeta.
\end{eqnarray}

The Carter constant plays a crucial role in the analysis of particle motion around a black hole, as it represents a conserved quantity arising from the separation of variables in the Hamilton-Jacobi equation, reflecting the hidden symmetry of the spacetime and its influence on geodesic motion. The Carter constant governs the polar ($\theta$-direction) dynamics of particles, depending on both the polar angular momentum and the orbital inclination relative to the black hole's equatorial plane. In astrophysical scenarios, the Carter constant is typically non-negative. This aligns with the assumption that the particle's motion is constrained by the geometry of the black hole's spacetime, with the tilt angle $\zeta$ providing a means to quantify this deviation from the equatorial plane. Therefore, in this study, we restrict our analysis to ${\cal K} \geq 0$.

We focus on circular orbits, which satisfy:
\begin{eqnarray}
	{\cal R}(r)={\cal R}'(r)=0,
\end{eqnarray}
where the prime denotes the derivative with respect to the radial coordinate $r$.

To analyze these orbits, we examine three conserved quantities: energy $E$, angular momentum $L$, and the Carter constant $\mathcal{K}$. These arise from spacetime symmetries and play a central role in determining particle dynamics near black holes. The energy $E$ is associated with the timelike Killing vector and controls motion in time, the angular momentum $L$ corresponds to the azimuthal Killing vector and governs motion in the azimuthal direction, and the Carter constant $\mathcal{K}$ emerges from the separability of the Hamilton–Jacobi equation and encodes information about polar (latitudinal) motion. The dependence of these quantities on the radial coordinate $r$ provides insight into orbital stability and the nature of relativistic precession.

\begin{figure}
\center{\subfigure[]{\label{rEnergy1a}
\includegraphics[width=5cm]{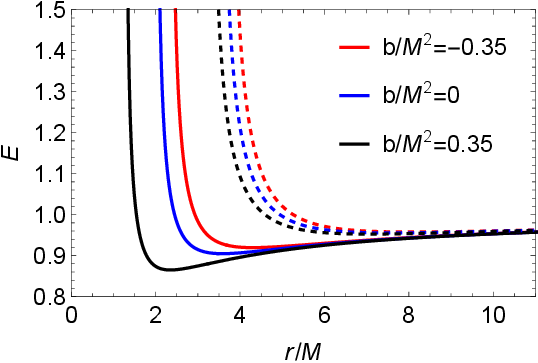}}
\subfigure[]{\label{rMomentum1b}
\includegraphics[width=5cm]{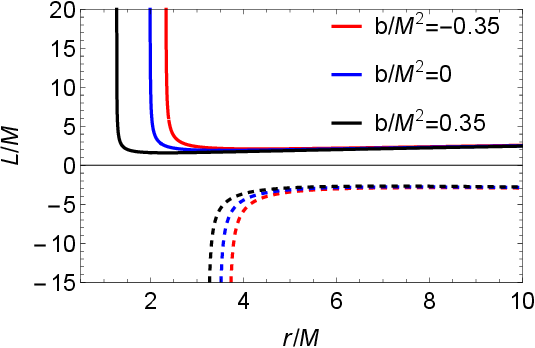}}
\subfigure[]{\label{rCarter1c}
\includegraphics[width=5cm]{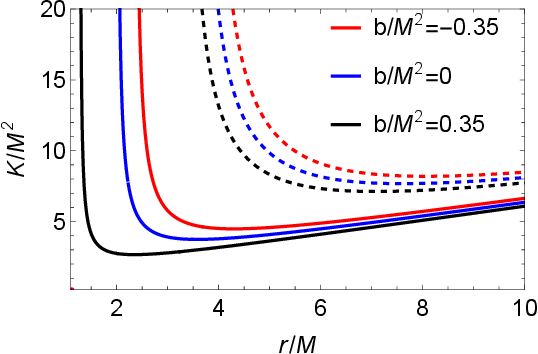}}\\
\subfigure[]{\label{rEnergy1d}
\includegraphics[width=5cm]{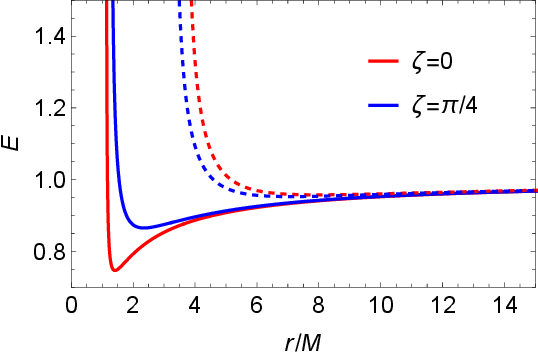}}
\subfigure[]{\label{rMomentum1e}
\includegraphics[width=5cm]{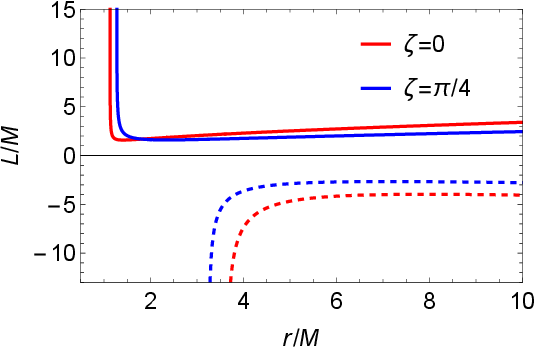}}
\subfigure[]{\label{rCarter1f}
\includegraphics[width=5cm]{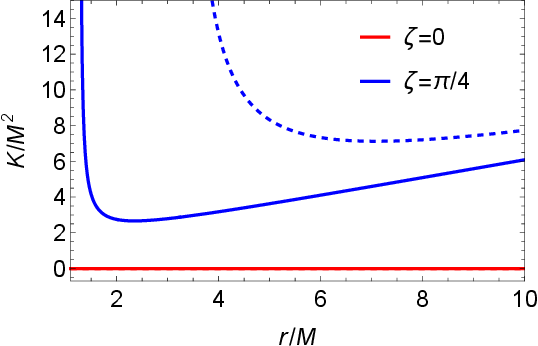}}}
\caption{The conserved quantities—energy $E$, angular momentum $L$ and Carter constant ${\cal K}$—for massive particles in circular orbits around a rotating braneworld black hole with spin parameter $a/M=0.8$. Top panels show variation with radial coordinate $r$ at fixed tilt angle $\zeta=\frac{\pi}{4}$ for different tidal charges $b$. Bottom panels display dependence on $r$ at fixed $b/M^2=0.35$ for various tilt angles $\zeta$. Solid and dashed curves represent prograde and retrograde orbits respectively.}\label{conserved}
\end{figure}

We now investigate the behavior of $E$, $L$, and $\mathcal{K}$ for circular orbits around a rotating braneworld black hole. Fig. \ref{conserved} displays their variations with radius of the spherical orbit $r$, for two distinct scenarios: (i) varying tidal charge $b$ (top panels) and (ii) varying tilt angle $\zeta$ (bottom panels). Throughout the analysis, we take the spin parameter $a/M = 0.8$. The solid curves represent the prograde orbits, while the dashed curves correspond to the retrograde ones.

In the top panels of Fig. \ref{conserved}, we examine the effect of tidal charge $b$ on the conserved quantities for $\zeta=\frac{\pi}{4}$. In Fig. \ref{rEnergy1a}, the energy $E$ decreases sharply as the orbital radius $r$ increases, then gradually flattens out. For both prograde and retrograde orbits, the energy decreases with $b$, indicating that positive tidal charges enhance gravitational binding, thereby reducing the energy required to sustain circular motion. Additionally, retrograde orbits always require more energy than their prograde counterparts at the same radius. In Fig. \ref{rMomentum1b}, the angular momentum $L$ shows a non-monotonic trend: it initially decreases steeply with increasing $r$, then gradually increases. This sharp initial drop near the black hole reflects the strong gravitational pull, which demands greater angular momentum for particles to remain in circular orbits. As the radius increases, the gravitational field weakens and the required angular momentum stabilizes. Larger values of $b$ correspond to lower angular momentum at a given radius, indicating that the tidal charge reduces the centrifugal support needed for orbital balance. The sensitivity to $b$ is more pronounced in prograde orbits. In Fig. \ref{rCarter1c}, the Carter constant ${\cal K}$ decreases rapidly at small radii and then increases slowly at larger distances. Similar to energy and angular momentum, ${\cal K}$ decreases with increasing $b$, reflecting a more confined polar motion due to stronger gravitational attraction. Retrograde orbits again exhibit higher values of ${\cal K}$ than prograde ones, consistent with their broader angular oscillation.

The bottom panels of Fig. \ref{conserved} illustrate how the tilt angle $\zeta$ affects the conserved quantities when the tidal charge is fixed at $b/M^2=0.35$. In Fig. \ref{rEnergy1d}, increasing the tilt angle results in a more gradual decrease of energy $E$ with radius, implying that particles on tilted spherical orbits require more energy to maintain circular motion. Equatorial orbits are energetically more favorable, indicating greater stability in the absence of inclination. In Fig. \ref{rMomentum1e}, for prograde motion, spherical orbits demand greater angular momentum near the black hole, but require less at larger radii compared to equatorial orbits. For retrograde orbits, angular momentum consistently decreases with increasing tilt angle, and the effect is more pronounced than in the prograde case. This asymmetry highlights the complex interplay between orbital inclination and rotational frame-dragging. In Fig. \ref{rCarter1f}, the Carter constant increases substantially with the tilt angle, especially in retrograde motion. It indicates that the tilt angle of the orbital plane directly affects the Carter constant, with spherical orbits requiring a larger value of ${\cal K}$ to remain the state. This is expected, as ${\cal K}$ quantifies deviations from equatorial symmetry. 

In summary, the conserved quantities—energy $E$, angular momentum $L$, and Carter constant ${\cal K}$—demonstrate clear and systematic dependence on both the orbital radius and the spacetime parameters $b$ (tidal charge) and $\zeta$ (tilt angle). Increasing the tidal charge $b$ leads to a reduction in all three conserved quantities, indicating enhanced gravitational binding in the presence of positive braneworld corrections. In contrast, increasing the tilt angle $\zeta$ results in higher values of $E$ and ${\cal K}$ due to stronger polar motion, and alters $L$ in a direction-dependent manner—increasing it for prograde orbits at small radii and decreasing it for retrograde orbits. These results offer essential insights into the orbital structure and provide a foundation for analyzing precession and stability in braneworld black hole spacetimes.

\subsection{Non-equatorial innermost stable spherical orbits} 

The ISSO is a fundamental property of black hole spacetime, marking the boundary between stable orbital motion and dynamical infall. For non-equatorial orbits around a rotating braneworld black hole, the ISSO radius $r_{\rm ISSO}$ depends sensitively on both the tidal charge $b$ and the orbital tilt angle $\zeta$.

The ISSO is determined by requiring that the radial potential $\mathcal{R}(r)$ satisfies the following conditions simultaneously:
\begin{eqnarray}
	\mathcal{R}(r)=\mathcal{R}'(r)=\mathcal{R}''(r)=0.
\end{eqnarray}

\begin{figure}
\center{\subfigure[]{\label{rissoequator}
\includegraphics[width=5cm]{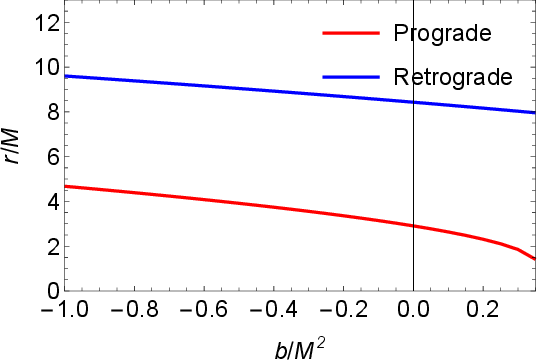}}
\subfigure[]{\label{rissopole}
\includegraphics[width=5cm]{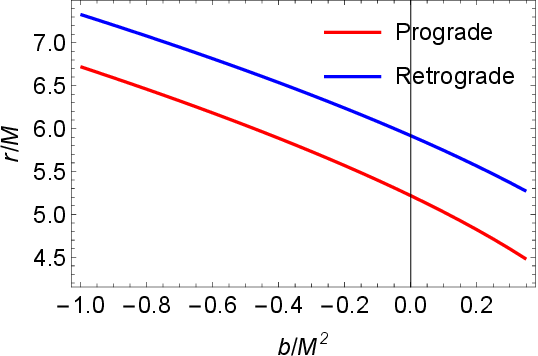}}}
\caption{ISSO radius as a function of tidal charge $b$ for (a) nearly equatorial orbits ($\zeta=0.01$) and (b) highly tilted orbits ($\zeta=1.45$), with fixed spin parameter $a/M=0.8$. Red and blue curves denote prograde and retrograde orbits respectively. Note different vertical scales between panels reflecting inclination dependence.}\label{risso}
\end{figure}

Fig. \ref{risso} presents a systematic analysis of ISSO radii across the physically allowed parameter space, highlighting how extra-dimensional effects encoded in $b$ interact with the geometric inclination of the orbit governed by $\zeta$. Two representative cases are shown: nearly equatorial orbits with $\zeta = 0.01$ (Fig. \ref{rissoequator}) and highly tilted orbits with $\zeta = 1.45$ (Fig. \ref{rissopole}). The spin is fixed at $a/M=0.8$.

In Fig. \ref{rissoequator}, we examine near-equatorial orbits. For prograde motion (red curves), the ISSO radius decreases rapidly as $b/M^2$ increases from -1 to 0.35, with the radius shrinking by nearly 70\%. This indicates that positive tidal charge allows stable circular motion closer to the black hole, as the gravitational pull becomes stronger due to higher spacetime curvature. In contrast, retrograde orbits (blue curves) are less sensitive to the changes in $b$, exhibiting only about a 16.7\% reduction in the ISSO radius across the same $b$ range.

In Fig. \ref{rissopole}, the analysis shifts to highly tilted orbits. The general trend remains—the ISSO radius decreases with increasing $b$. While the radius of prograde orbits with high inclination is consistently larger than that of their equatorial counterparts for any given $b$. Interestingly, for retrograde orbits, the opposite holds: tilted ISSOs lie slightly below those of equatorial motion. While the dependence on $b$ is still evident in prograde cases, it is noticeably weakened compared to equatorial orbits. Moreover, as $\zeta$ approaches $\frac{\pi}{2}$, the distinction between prograde and retrograde ISSOs diminishes; at $\zeta=\frac{\pi}{2}$, the two trajectories are expected to coincide due to the axial symmetry of the system.

The physical interpretation of these results becomes clear when examining how the tidal charge $b$ and the orbital tilt angle $\zeta$ influence the underlying spacetime geometry. Positive tidal charges ($b>0$) effectively strengthen the gravitational field near the black hole by introducing additional curvature sourced from the higher-dimensional bulk. This enhanced binding energy allows particles—particularly those on prograde orbits—to maintain stable circular motion at smaller radii. In contrast, negative values of $b$ induce a repulsive tidal component that weakens the effective gravitational pull, thereby pushing the ISSO outward.

The inclination angle $\zeta$ governs how strongly the orbital plane couples to the black hole's frame-dragging effect. Orbits near the equatorial plane ($\zeta \approx 0$) experience maximal rotational coupling, which enhances the influence of the black hole’s spin and tidal charge on the orbital dynamics. Conversely, orbits approaching the polar limit ($\zeta \approx \pi/2$) decouple from these effects, resulting in reduced sensitivity to both $a$ and $b$. Consequently, the combined modulation by $b$ and $\zeta$ defines a rich parameter space in which prograde and retrograde motions exhibit distinct behaviors. 

These discoveries have significant implications for astrophysical observations. First, since the ISSO sets the inner edge of the accretion disk, its radius directly affects the high-energy cutoff in the thermal emission spectrum, offering a sensitive probe of the underlying spacetime geometry. Second, QPOs observed in microquasars or active galactic nuclei may exhibit frequency shifts that correlate with variations in the tidal charge. Third, gravitational wave signals from extreme mass-ratio inspirals (EMRIs) could encode subtle signatures of braneworld effects in their late-stage inspiral trajectories.

In conclusion, precise measurements of ISSO radii—via electromagnetic observation or gravitational wave detection—offer a promising avenue for probing braneworld gravity. In particular, the strong sensitivity of equatorial prograde ISSO to the tidal charge (as shown in Fig. \ref{risso}) makes them especially valuable targets for testing extra-dimensional effects. Future high-precision instruments, such as the Next Generation Event Horizon Telescope (ngEHT) and LISA, may provide the necessary resolution to detect or constrain these high-dimensional corrections, thus providing critical tests of alternative gravity models beyond GR.

\section{Epicyclic frequencies and orbital precession}
\label{sec:particleprecession}

Test particle motion around black holes is not restricted to exact circular orbits. When slightly perturbed, particles undergo small oscillations in both radial and vertical directions, characterized by distinct epicyclic frequencies. These frequencies reflect the curvature and rotational structure of spacetime and play a central role in understanding orbital dynamics, stability, and observable relativistic precession phenomena.

In the previous section, we have investigated the conserved quantities and ISSO behavior of test particles on circular orbits. We now extend this analysis by considering small perturbations to these orbits and derive the corresponding epicyclic frequencies: the radial frequency $\Omega_r$, the vertical or polar frequency $\Omega_\theta$, and the azimuthal orbital frequency $\Omega_\phi$. The differences among these frequencies give rise to two key precession effects: nodal precession and periastron precession, which provide observational signatures of GR and its potential modifications in braneworld scenarios.

Nodal precession refers to the gradual rotation of the orbital plane around the spin axis of the black hole. It is caused by the discrepancy between the azimuthal and vertical oscillation frequencies, and directly encodes the effect of frame-dragging. Periastron precession, on the other hand, describes the rotation of the orbit’s point of closest approach (periastron) within the orbital plane. It arises from the difference between the azimuthal and radial frequencies and reflects spacetime curvature effects. Both types of precession are sensitive to the black hole's spin parameter $a$ and the tidal charge $b$.

In the context of braneworld gravity, the tidal charge introduces significant modifications to the epicyclic structure, offering a potential observational handle on extra dimensions. Compared to standard Kerr black holes, the presence of $b$ alters the geometry of the effective potential and frame-dragging behavior, leading to measurable shifts in precession frequencies.

In this section, we will investigate the precession phenomena experienced by test particles on nearly circular orbits around a rotating braneworld black hole. While previous analyses focused on the properties of stable equatorial orbits, we now explore how small perturbations to these orbits give rise to rich dynamical behavior characterized by distinct oscillation frequencies. We begin by deriving the fundamental frequencies from the effective potential and metric components, and then analyze how they vary with the black hole parameters and orbital radius.

\subsection{Nodal precession and frame-dragging effects}

Nodal precession is a relativistic phenomenon arising from frame-dragging in curved spacetime, characterized by the gradual rotation of a particle’s orbital plane around the black hole’s spin axis. It occurs when an orbit deviates slightly from the equatorial plane, introducing vertical oscillations governed by the frequency $\Omega_\theta$, which interact with the primary azimuthal motion at frequency $\Omega_\phi$. First predicted by Lense and Thirring in 1918 \cite{Lense:1918zz}, this effect is a direct consequence of rotating masses dragging the surrounding spacetime, analogous to a spinning object stirring a viscous fluid.

The physical origin of nodal precession can be understood from three perspectives. First, the black hole’s spin generates a gravitomagnetic field, which exerts a torque on the particle’s orbital angular momentum vector. Second, vertical perturbations lead to harmonic oscillations perpendicular to the equatorial plane, with a characteristic frequency $\Omega_\theta$. Third, a mismatch between the orbital period $2\pi/\Omega_\phi$ and the vertical oscillation period $2\pi/\Omega_\theta$ leads to a cumulative shift of the orbital plane over time.

In the braneworld scenario, the presence of a tidal charge $b$ modifies the geometry of spacetime and thus alters the precession dynamics. Specifically, the metric component $g_{t\phi}$, responsible for frame-dragging, acquires additional $b$-dependent terms, modifying the gravitomagnetic potential. The effective potential $V_{\text{eff}}$ also develops changes in curvature in the vertical direction. Moreover, the relation between the coordinate radius and proper time is modified, which in turn affects the timing of orbital motion.

We begin the mathematical analysis by introducing the orbital angular frequency $\Omega_\phi$ for a test particle on a circular equatorial orbit in the rotating braneworld spacetime:
\begin{eqnarray}
	\Omega_{\phi}=\frac{-g_{t\phi,r}+\sqrt{(g_{t\phi,r})^2-g_{tt,r} g_{\phi\phi,r}}}{g_{\phi\phi,r}},
	\label{eq:omega_phi}
\end{eqnarray}
where a comma denotes the derivative with respect to the corresponding coordinate. When small perturbations are introduced, the particle undergoes vertical oscillations with an associated epicyclic frequency. The vertical epicyclic frequency $\Omega_\theta$ can be derived from second-order perturbation theory applied to the $\theta$-equation of motion \cite{Ryan:1995wh,Doneva:2014uma}:
\begin{eqnarray}
	\Omega_\theta=\sqrt{\frac{1}{2g_{\theta\theta}}\left[X^2 \partial_\theta^2 \left(\frac{g_{\phi\phi}}{g_{tt}g_{\phi\phi}-g_{t\phi}^2}\right)-2X Y\partial_\theta^2 \left(\frac{g_{t\phi}}{g_{tt}g_{\phi\phi}-g_{t\phi}^2}\right)+Y^2 \partial_\theta^2 \left(\frac{g_{tt}}{g_{tt}g_{\phi\phi}-g_{t\phi}^2}\right)\right]},
\end{eqnarray}
where
\begin{eqnarray}
	X=g_{tt}+g_{t\phi}\Omega_{\phi}, Y=g_{t\phi}+g_{\phi\phi} \Omega_{\phi}.
\end{eqnarray}

The nodal precession frequency is defined as the difference between the azimuthal and vertical frequencies:
\begin{eqnarray}
	\Omega_{\text{nod}}=\Omega_\phi-\Omega_\theta\approx \frac{2aM}{r^{3}}-\frac{3a^{2}M^{1/2}}{2}\left(\frac{1}{r}\right)^{7/2}-\frac{ab}{r^{4}}+\mathcal{O}\left[\frac{1}{r}\right]^{9/2}.
	\label{eq:nodal_expansion}
\end{eqnarray}
This expansion reveals several key features. The leading term $2aM/r^3$ corresponds to the standard Lense-Thirring prediction in GR. The subsequent terms represent higher-order relativistic corrections and braneworld effects. In particular, the $b$-dependent term introduces a distinct radial scaling that alters both the magnitude and radial decay of the precession. A positive tidal charge $b$ enhances the precession rate at all radii, while a negative $b$ suppresses it. However, the impact of $b$ becomes negligible at large distances; specifically, beyond $r \gtrsim 20M$, the nodal precession frequency falls below detectable levels for current or near-future instrumentation.

\begin{figure}
\center{\subfigure[]{\label{bomeganod}
\includegraphics[width=5cm]{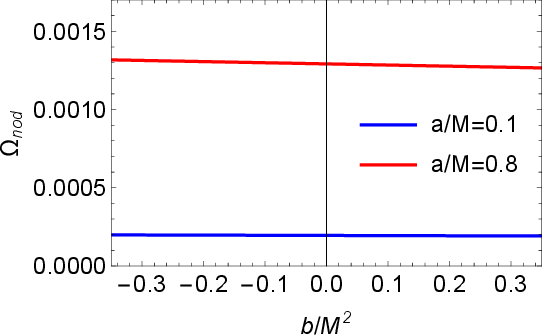}}
\subfigure[]{\label{aomeganod}
\includegraphics[width=5cm]{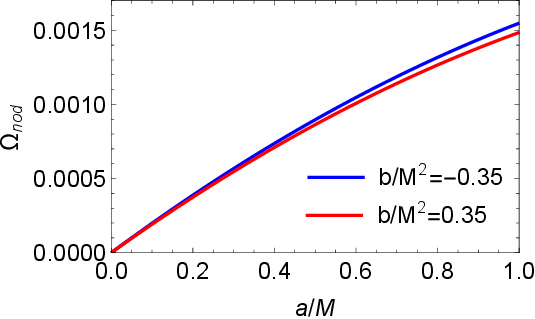}}
\subfigure[]{\label{romeganod}
\includegraphics[width=5cm]{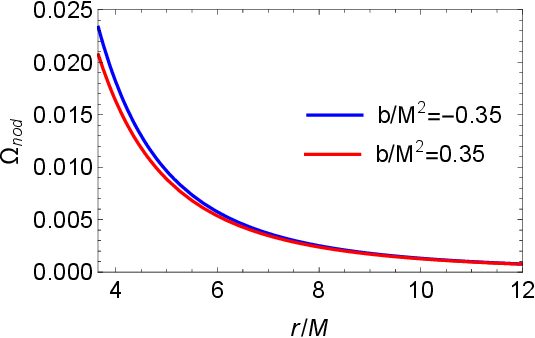}}}
\caption{Nodal precession frequency $\Omega_{\text{nod}}$ as a function of: (a) tidal charge $b$ at fixed $r/M = 10$ for spins $a/M = 0.1$ and $0.8$; (b) spin parameter $a$ at $r/M = 10$ for $b/M^2 = \pm 0.35$; (c) radial coordinate $r$ for $a/M = 0.8$ and $b/M^2 = \pm 0.35$.}\label{omeganod}
\end{figure}

Fig. \ref{omeganod} illustrates how the nodal precession frequency depends on the tidal charge $b$, black hole spin $a$, and orbital radius $r$. In Fig. \ref{bomeganod}, we fix the orbital radius at $r/M=10$ and compare the variation of $\Omega_{\text{nod}}$ with respect to $b$ for two values of the spin parameter ($a/M=0.1$ and 0.8). For both spin values, the precession frequency decreases as $b$ increases, with higher spin leading to stronger frame-dragging. In Fig. \ref{aomeganod}, we plot $\Omega_{\text{nod}}$ as a function of $a$ for fixed $b$. The relationship is nearly linear, and the slope becomes flatter as $b$ increases, consistent with the interpretation that positive tidal charge suppresses frame-dragging effects. In Fig. \ref{romeganod}, we explore the radial dependence, showing that $\Omega_{\text{nod}}$ decays rapidly with radius, following the typical $r^{-3}$ behavior of gravitomagnetic effects.

The physical interpretation of these results becomes clear when considering how the tidal charge $b$ affects the spacetime geometry. In all cases, $\Omega_{\text{nod}}$ decreases with $b$, consistent with the interpretation that a positive tidal charge enhances gravitational binding and suppresses frame-dragging. The spin parameter $a$ directly controls the frame-dragging effect, with higher spins producing stronger precession. Additionally, the radial dependence follows the typical $r^{-3}$ scaling of frame-dragging effects in GR, modified by the braneworld corrections.

From an observational standpoint, nodal precession manifests in a variety of astrophysical contexts. It influences the precession of tilted accretion disks, the modulation of QPOs in X-ray binaries, and the motion of stars orbiting supermassive black holes. In gravitational wave astronomy, the phase evolution of EMRIs encodes the nodal precession rate. Deviations from the Kerr baseline could provide evidence for the presence of extra-dimensional corrections. Future missions such as LISA may have the precision necessary to detect such signatures.

\subsection{Periastron precession and spacetime curvature}

Periastron precession is a relativistic effect that causes the point of closest approach in an orbit to advance with each revolution. It results from the difference between the orbital angular frequency $\Omega_\phi$ and the radial epicyclic frequency $\Omega_r$. In Newtonian gravity, bound orbits are closed ellipses. However, in GR and its extensions, curved spacetime leads to open orbits that precess over time. This periastron shift is particularly significant in strong-field regions near compact objects, making it a valuable probe of gravitational dynamics. 

The mathematical description of periastron precession in the braneworld scenario requires careful analysis of the radial epicyclic frequency $\Omega_r$. For a test particle on a slightly eccentric orbit in the vicinity of the equatorial plane, the radial motion can be described as a perturbation around a stable circular orbit \cite{Ryan:1995wh,Doneva:2014uma}:
\begin{eqnarray}
	\Omega_r=\sqrt{\frac{1}{2g_{rr}}\left[X^2\partial_r^2 \left(\frac{g_{\phi\phi}}{g_{tt}g_{\phi\phi}-g_{t\phi}^2}\right)-2XY \partial_r^2 \left(\frac{g_{t\phi}}{g_{tt}g_{\phi\phi}-g_{t\phi}^2}\right)+Y^2 \partial_r^2 \left(\frac{g_{tt}}{g_{tt}g_{\phi\phi}-g_{t\phi}^2}\right)\right]}.
\end{eqnarray}
The periastron precession frequency then follows as:
\begin{eqnarray}
	\Omega_{\text{per}}=\Omega_\phi-\Omega_r\approx 
	\frac{M^{3/2}}{r^{5/2}}\left(3-\frac{b}{2M^2}-\frac{4a}{r^{1/2} M^{1/2}}\right)+\mathcal{O}\left[\frac{1}{r}\right]^{7/2}.
	\label{eq:peri_approx}
\end{eqnarray}

The distinctive behavior of periastron precession in braneworld spacetimes becomes evident from the structure of the expansion. The leading-order term, $3M^{3/2}/r^{5/2}$, matches the Schwarzschild prediction. However, higher-order terms reveal how spin and extra-dimensional corrections jointly influence the precession dynamics. The tidal charge $b$ enters the expression through a characteristic modification of the radial dependence, and its sign determines whether precession is suppressed ($b>0$) or enhanced ($b<0$) relative to the four-dimensional Kerr case. This sign-dependent effect leaves a unique imprint that can distinguish braneworld corrections from other strong-field deviations.

The physical origin of this behavior lies in the way the bulk geometry deforms the intrinsic curvature of the brane. The tidal charge modifies the metric component $g_{rr}$, effectively rescaling the proper radial distance near the black hole. This geometric distortion alters the balance between gravitational attraction and centrifugal repulsion in the effective potential. A positive $b$ increases the curvature of the potential well, strengthening the restoring force during radial oscillations. As a result, particles complete epicyclic motion more quickly (larger $\Omega_r$), reducing the phase difference between the orbital and radial periods, and thereby lowering the periastron precession rate ($\Omega_{\text{per}}$).

\begin{figure}
\center{\subfigure[]{\label{bomegaper}
\includegraphics[width=5cm]{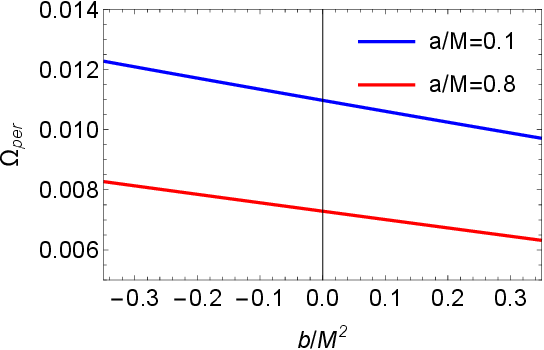}}
\subfigure[]{\label{aomegaper}
\includegraphics[width=5cm]{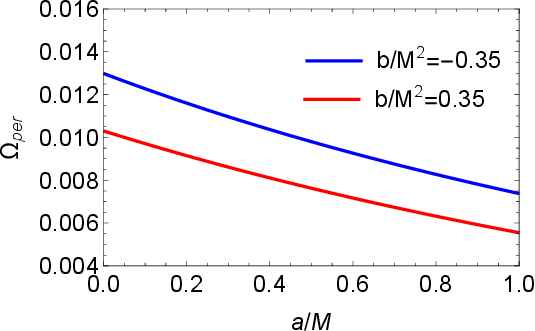}}
\subfigure[]{\label{romegaper}
\includegraphics[width=5cm]{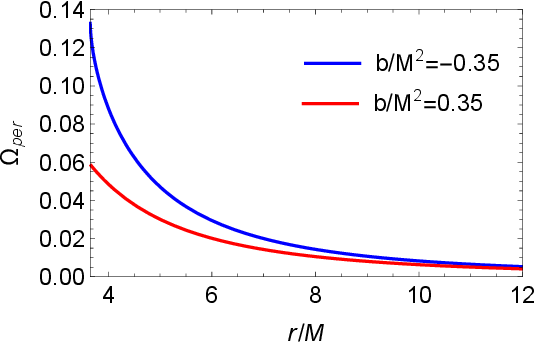}}}
\caption{Periastron precession frequency $\Omega_{\text{per}}$ as a function of: (a) tidal charge $b$ at fixed $r/M = 10$ for spins $a/M = 0.1$ and $0.8$; (b) spin parameter $a$ at $r/M = 10$ for $b/M^2 = \pm 0.35$; (c) radial coordinate $r$ for $a/M = 0.8$ and $b/M^2 = \pm 0.35$.}\label{omegaper}
\end{figure}

Fig. \ref{omegaper} illustrates the behavior of $\Omega_{\text{per}}$ under the same three configurations as in Fig. \ref{omeganod}. In Fig. \ref{bomegaper}, we fix $r/M=10$ and vary $b$, observing that $\Omega_{\text{per}}$ also decreases with $b$. It demonstrates how positive tidal charges suppress periastron precession, while negative charges enhance it. Fig. \ref{aomegaper} confirms the strong dependence of $\Omega_{\text{per}}$ on the black hole spin $a$. Fig. \ref{romegaper} displays the characteristic $r^{-5/2}$ radial scaling, modified by braneworld corrections that become particularly pronounced near the ISSO.

The different behavior of $\Omega_{\text{nod}}$ and $\Omega_{\text{per}}$ reflects the different mechanisms behind them: frame-dragging versus spacetime curvature. Their distinct dependencies on $b$, $a$, and $r$ offer complementary avenues for probing the geometry of the braneworld black hole.

In this section, we have analyzed the precessional motion of test particles in circular orbits around rotating braneworld black holes. Small perturbations to equatorial orbits give rise to radial and vertical epicyclic oscillations, characterized by the frequencies $\Omega_r$ and $\Omega_\theta$, respectively. These, in turn, lead to the emergence of two types of relativistic precession: nodal precession driven by frame-dragging, and periastron precession induced by spacetime curvature. Our results show that the tidal charge $b$ alters both types of precession, and that their sensitivity to spin $a$ and orbital radius $r$ varies in distinct ways. These effects provide a multifaceted diagnostic for testing the braneworld scenario through future electromagnetic and gravitational wave observations.

\section{Lense-Thirring precession of test gyroscope}
\label{sec:precessionlt}

In this section, we discuss the Lense-Thirring effect for a stationary gyroscope in the spacetime of a rotating braneworld black hole. The Lense-Thirring effect, also known as frame-dragging, describes how the rotation of a massive object twists the surrounding spacetime, causing the spin axis of a gyroscope to precess. This effect becomes especially pronounced near the event horizon of a rotating black hole.

We consider a gyroscope that remains stationary relative to a distant observer, implying that its orbital angular momentum vanishes ($L=0$) while its spatial position remains unchanged. In the braneworld scenario, the Lense-Thirring effect on such a stationary gyroscope is influenced by the tidal charge $b$, thereby modifying the precession rates compared to those of the standard Kerr black hole.

The precession frequency $\Omega_{\rm LT}$ for a stationary gyroscope is given by \cite{Chakraborty:2013naa}: 
\begin{eqnarray}
	\Omega_{\texttt{LT}}=\frac{1}{2\sqrt{-g}}\bigg(\sqrt{g_{\theta \theta}}\Big(g_{t\phi,r}-\frac{g_{t\phi}}{g_{tt}}g_{tt,r}\Big) \hat{\theta}-\sqrt{g_{rr}}\Big(g_{t\phi,\theta}-\frac{g_{t\phi}}{g_{tt}} g_{tt,\theta}\Big)\hat{r}\bigg),
\end{eqnarray}
where the metric components are evaluated at the fixed position of the stationary gyroscope. This expression capturesc how spacetime dragging induces rotation of the gyroscope’s spin axis. The first term $\Big( g_{t\phi,r}-\frac{g_{t\phi}}{g_{tt}} g_{tt,r} \Big)$ accounts for radial variations in frame-dragging, while the second term $\Big( g_{t\phi,\theta}-\frac{g_{t\phi}}{g_{tt}} g_{tt,\theta} \Big)$ captures polar variations. The derivatives $\partial_{\theta}$ and $\partial_r$ are used to measure how the angle of the orbit changes in response to the spacetime curvature in these directions. By solving these equations and incorporating the tidal charge $b$ and the black hole's spin $a$, we can calculate the precession frequency $\Omega_{\texttt{LT}}$. Additionally, we also take into account the effects arising from the orientation angle. For a stationary gyroscope case, we analyze $\Omega_{\rm LT}$ as functions of these parameters, with the radial coordinate fixed at $r/M=10$. This allows us to investigate how different black hole parameters affect the precession of a gyroscope that's maintaining fixed position against the black hole's gravity.

\begin{figure}
\center{\subfigure[]{\label{bomegalt}
\includegraphics[width=5cm]{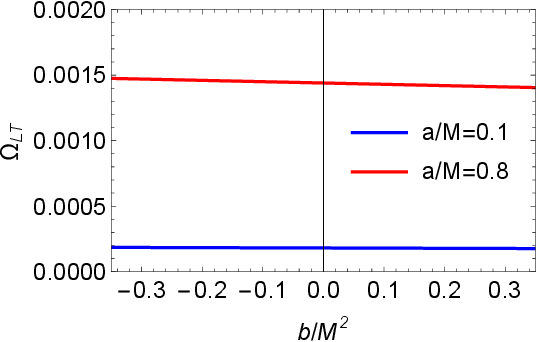}}
\subfigure[]{\label{aomegalt}
\includegraphics[width=5cm]{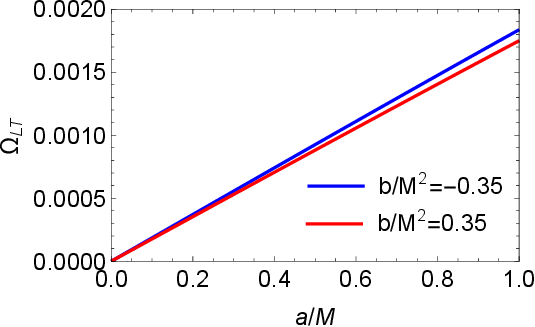}}
\subfigure[]{\label{zetaomegalt}
\includegraphics[width=5cm]{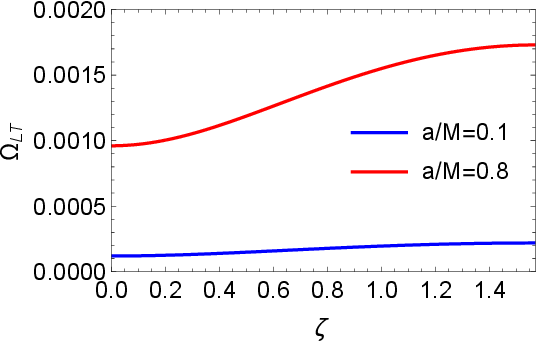}}}
\caption{Lense-Thirring precession frequency $\Omega_{\texttt{LT}}$ as a function of: (a) tidal charge $b$ with $\zeta=\frac{\pi}{4}$; (b) spin parameter $a$ with $\zeta=\frac{\pi}{4}$; (c) tilt angle $\zeta$ at $b/M^2=0.35$. We set $r/M = 10$.}\label{omegalt}
\end{figure}

We begin by examining the dependence of the Lense-Thirring frequency on the tidal charge $b$. As shown in Fig. \ref{bomegalt}, we compare results for two spin parameters: $a/M=0.1$ and $a/M=0.8$. The precession frequency $\Omega_{\rm LT}$ shows distinct trends as $b$ varies from negative to positive values. For $a/M=0.1$, $\Omega_{\rm LT}$ remains nearly constant ($\Omega_{\texttt{LT}}|_{b=-0.35M^2}=0.000186$ and $\Omega_{\texttt{LT}}|_{b=0.35M^2}=0.000177$, corresponding to a negligible rate of change of -0.001261\%). In contrast, for $a/M=0.8$, the frequency decreases more significantly ($\Omega_{\texttt{LT}}|_{b=-0.35M^2}=0.00147$ and $\Omega_{\texttt{LT}}|_{b=0.35M^2}=0.00140$, with a rate of change of -0.010015\%). This demonstrates that the tidal charge’s influence on $\Omega_{\rm LT}$ grows with black hole spin: larger spins ($a/M=0.8$) exhibit stronger sensitivity to $b$, while smaller spins ($a/M=0.1$) show minimal variation. Physically, this occurs because positive tidal charges enhance the gravitational attraction near the black hole, thereby suppressing frame-dragging effect and reducing $\Omega_{\rm LT}$. Conversely, negative tidal charges weaken the gravitational pull, amplifying frame-dragging and increasing $\Omega_{\rm LT}$. In summary, these results highlight the tidal charge’s critical role in modifying spacetime geometry and precession dynamics.

Next, we examine the relationship between the Lense-Thirring frequency and the black hole's spin parameter $a$, by considering two tidal charges: $b/M^2=-0.35$ and $b/M^2=0.35$. Fig. \ref{aomegalt} illustrates that $\Omega_{\rm LT}$ monotonically increases with $a$. This result suggests that as the spin of black hole increases, the frame-dragging effect becomes stronger, leading to faster precession of particle orbits. The spin parameter $a$ directly influences the black hole's angular momentum, thereby determining the extent to which spacetime is dragged along with the black hole's rotation. This relationship further emphasizes the importance of the black hole's spin in determining the precession behavior of spherical orbits. Additionally, negative tidal charge $b$ exhibits greater sensitivity than positive $b$. This underscores the interdependent roles of spin and tidal charge in modifying precession dynamics.

Finally, we investigate the impact of the orbital tilt angle $\zeta$ on the Lense-Thirring frequency, considering two spin parameters ($a/M=0.1$ and $a/M=0.8$) with a fixed tidal charge $b/M^2=0.35$. Fig. \ref{zetaomegalt} demonstrates that $\Omega_{\rm LT}$ grows with $\zeta$, showing a much stronger dependence for rapidly spinning black holes ($a/M=0.8$) than for slowly rotating ones ($a/M=0.1$). This behavior occurs because greater misalignment between the gyroscope's spin axis and the black hole's rotation axis (larger $\zeta$) enhances frame-dragging effects, thereby increasing $\Omega_{\rm LT}$. The effect intensifies with higher spin parameters, revealing how strongly the black hole's rotation couples to tilted spherical orbital geometries. These results collectively demonstrate the intricate relationship between orbital configuration and rotational dynamics in braneworld black holes.

Regarding Fig. \ref{bomegalt}, we address an apparent paradox: while positive tidal charges $b>0$ enhance the gravitational field, they simultaneously weaken the Lense-Thirring precession. This behavior stems from distinct ways the tidal charge affects spacetime geometry versus frame-dragging dynamics. Positive tidal charge indeed enhances gravitational curvature by modifying both the potential and metric components, particularly near the horizon. However, this enhancement does not necessarily imply a strengthening of the precession effect in all cases. In fact, positive tidal charges may change the spacetime structure surrounding the gyroscope, thereby weakening the precession effect. This can be interpreted in two ways: 

(1) Modification of frame-dragging: A positive tidal charge alters the way the black hole’s spin couples to the gyroscope’s orientation. This change reduces the influence of rotation on the local inertial frames, leading to slower or more stable precession.

(2) Change in spacetime geometry: The tidal charge reshapes the gravitational field near the black hole, modifying how the spin vector of the gyroscope evolves. Although the curvature increases, the specific structure of the metric components reduces the effect of rotation on precession. Effectively, a positive $b$ decreases the rotational influence of the black hole, diminishing its impact on nearby gyroscopes. Hence, the simultaneous enhancement of gravity and suppression of precession by positive tidal charge reflects distinct aspects of the tidal charge’s role in spacetime structure.

In summary, we have studied the Lense-Thirring precession of stationary gyroscopes in the spacetime of rotating braneworld black holes. The precession frequency $\Omega_{\texttt{LT}}$ is found to depend sensitively on the black hole’s spin $a$, tidal charge $b$, and the orientation angle $\zeta$ of the gyroscope’s spin axis. The frequency increases with spin and tilt angle, while it decreases with increasing positive tidal charge. These results deepen our understanding of frame-dragging in extra-dimensional gravity models and may inform precision tests using gyroscopic measurements near compact objects. Observational methods such as space-based gyroscope missions and gravitational wave detections offer potential avenues for probing braneworld effects and constraining the tidal charge.

\section{Summary}
\label{sec:summary}

We have presented a comprehensive study of orbital dynamics and precession effects in the spacetime of rotating braneworld black holes, exploring both test particle motion and stationary gyroscopes. Our analysis demonstrates how the braneworld tidal charge $b$ alters key relativistic phenomena, offering potential observational signatures of extra-dimensional gravity.

For test particle motion, we first analyzed the three conserved quantities—energy $E$, angular momentum $L$, and Carter constant $\mathcal{K}$—for spherical orbits. We found that the tidal charge $b$ exhibits an inverse correlation with all three quantities, while the tilt angle $\zeta$ enhances both $E$ and $\mathcal{K}$. The ISSO radius shows particularly notable behavior, decreasing significantly with positive $b$ for prograde orbits, while the retrograde case exhibits more modest changes.

The precession phenomena show clear braneworld signatures. Both nodal precession (from frame-dragging effect) and periastron precession (from spacetime curvature) display $b$-dependent modifications to their characteristic frequencies. Positive tidal charges suppress $\Omega_{\rm nod}$, whereas negative charges enhance it, with the strongest deviations appearing at small radii. Periastron precession $\Omega_{\rm per}$ follows a qualitatively similar trend, albeit with a distinct radial profile, providing complementary probes of the underlying spacetime structure.

For stationary gyroscopes, we computed the Lense-Thirring precession frequency $\Omega_{\rm LT}$ and examined its dependence on the tidal charge, black hole spin, and gyroscope orientation. A key result is that positive $b$ weakens the precession effect, despite amplifying the gravitational field strength—a consequence of modified frame-dragging due to the extra-dimensional contribution to the spacetime geometry. The orientation angle plays a critical role: larger misalignment angles result in stronger precession, especially for rapidly spinning black holes.

These theoretical results have direct implications for astrophysical observations. Shifts in the ISSO location can influence the inner edge of accretion disks and modify high-energy cutoffs in their emission spectra. QPOs in X-ray binaries may exhibit frequency patterns correlated with the tidal charge. In the Galactic Center, both stellar orbital dynamics and future precision gyroscope experiments could be used to test these effects. Additionally, gravitational wave signals from EMRIs are expected to encode precession signatures in their waveform phase evolution.

While current observations are not yet sensitive enough to place stringent constraints on the tidal charge, upcoming facilities such as LISA and the ngEHT may offer the necessary precision. Our work lays the theoretical foundation for interpreting these future measurements. In short, we have identified multiple distinct channels—ranging from orbital dynamics to gyroscopic precession—through which braneworld black holes imprint observable signatures, potentially opening a new window into extra-dimensional physics.

\section*{Acknowledgements}
This work was supported by the National Natural Science Foundation of China (Grants No. 12222302, No. 12475055, and No. 12447137), and National Key Research and Development Program of China (No. 2024YFC2207400).

\end{document}